\documentclass[twoside,11pt,reqno]{amsart}
\usepackage{a4wide}
\usepackage{amsmath}
\usepackage{amsfonts}
\usepackage{amssymb}
\usepackage{mathrsfs}
\usepackage{times}
\usepackage[arrow, matrix, curve]{xy}

\usepackage{pstricks,pst-node,pst-text,pst-3d}
\usepackage{graphicx}

\psset{framearc=0.0,framesep=0.0truecm,nodesep=3pt,unit=0.75truecm,%
arrowsize=4pt 2,arrowinset=0.4}

\newcounter{mycnt}

\def\!{\kern -0.15ex}

\def\piprod{\raisebox{0.2 ex}{${\scriptstyle \odot}$}\kern .2ex}
\def\ov{\overline}
\def\1{\ov{1}}
\def\2{\ov{2}}
\def\3{\ov{3}}

\begin{document}

\title[Strand symmetric model]
{Matrix group structure and Markov invariants in the strand symmetric phylogenetic substitution model} 

\author{Peter D Jarvis}
\address{P D Jarvis,  %
School of Mathematics and Physics, University of Tasmania, 
Private Bag 37, GPO, Hobart Tas 7001, Australia}
\email{Peter.Jarvis@utas.edu.au}
\author{Jeremy G Sumner}
\address{J G Sumner,  %
School of Mathematics and Physics, University of Tasmania, 
Private Bag 37, GPO, Hobart Tas 7001, Australia}
\email{Jeremy.Sumner@utas.edu.au}



\date{July 2013}

\begin{abstract}
We consider the continuous-time presentation of the strand symmetric phylogenetic substitution model (in which rate parameters are unchanged under nucleotide permutations given by Watson-Crick base conjugation). 
Algebraic analysis of the model's underlying structure as a matrix group leads to a change of basis where the rate generator matrix is given by a two-part block decomposition. 
We apply representation theoretic techniques and,  for any (fixed) number of phylogenetic taxa $L$ and polynomial degree $D$ of interest, provide the means to classify and enumerate the associated Markov invariants.
In particular, in the quadratic and cubic cases we prove there are precisely $\frac{1}{3}(3^L+(-1)^L)$ and $6^{L-1}$ linearly independent Markov invariants, respectively. 
Additionally, we give the explicit polynomial forms of the Markov invariants for (i) the quadratic case with any number of taxa $L$, and (ii) the cubic case in the special case of a three-taxa phylogenetic tree.  
We close by showing our results are of practical interest since the quadratic Markov invariants provide independent estimates of phylogenetic distances based on (i) substitution rates \emph{within} Watson-Crick conjugate pairs, and (ii) substitution rates \emph{across} conjugate base pairs.
\end{abstract}
\maketitle

\section{Introduction and motivation}
Recent years have seen rapid advances in the quantity and variety of molecular-based sequence data available for analysis and interpretation in terms of biological structure, function and evolution. Whole \emph{genome} datasets are increasingly accompanied by other types of `\emph{--omic}' data: \emph{transcriptome}, \emph{proteome}, \emph{metabolome}, amongst others. 
In turn, all of these modes of data representation require adequate mathematical model building in stochastic settings in order to capture the essential process systematics with parsimonious parametrizations.

Despite these ongoing challenges, the original brief of phylogenetics -- the use of quantitative, inter-species comparison data (in the modern context, molecular sequence data) to infer the evolutionary ancestry of species -- remains central.
It is still the contention that quality data, based on suitably aligned molecular sequences, should admit analysis via appropriate parametric probability models consistent with the neutral theory of evolution. The aim is a statement of taxonomic ancestry via an inferred phylogenetic tree, or perhaps a network representation which encapsulates unresolved ambiguities in the data. 
Under further assumptions about absolute mutation rates, parameter estimation then permits recovery of evolutionary divergence times (see \cite{felsenstein2004} for general background on phylogenetic methods).  

For nucleic acid base sequence data, the so-called general Markov model is in practice specialized, so that the key theoretical object -- an assumed $4\!\times\! 4$ stochastic matrix of base substitutions -- is not parametrized in the most general possible way.  A popular choice for maximum likelihood calculations is the general time-reversible (\texttt{GTR}) model \cite{tavare1986}; further constraints on the parameters lead to one of a number of other model types. Amongst these, we distinguish the ``group-based'' models (\cite{semple2003}, chapter 8), which allow for direct analytical treatments, using discrete Fourier or Hadamard inversion techniques \cite{hendy1989b,szekely1993}. 

The armoury of theoretical techniques has been further enriched with the advent of algebraically-inspired methods which seek to locate certain geometric structures, defined by the embedding of the models' parameter space into the multivariate probability spaces populated by the sequence data. 
Theoretical work around this approach is part of the relatively new field of ``algebraic statistics'' \cite{pachter2005}. 

Turning to computational approaches, although maximum likelihood optimization is powerful enough to allow full parameter recovery, in principle even for the general Markov model \cite{chang1996}, in practical implementations it is usual to work with specialized models. In \cite{sumner2011} we argued for the natural criterion of \emph{closure} (under matrix multiplication) as a guide to model choice in phylogenetics. In that work it was shown that \texttt{GTR} generically fails to be multiplicative closed, and our subsequent work with simulations showed how serious errors in phylogenetic estimation could potentially arise as a result \cite{sumner2012a}. 
Beyond the group-based models, we have studied a large class of closed models based on matrix Lie groups, the so-called Lie Markov models \cite{sumner2011}. 
In the continuous time context, these models have affiliated Lie algebras where the rate matrices are contained within an appropriate stochastic cone (see \cite{sumner2012c} for details). 

Of course, the general Markov model itself is by construction multiplicatively closed, and in related work \cite{sumner2008,sumner2009} we have exploited its matrix group structure to construct many new polynomials in the probability tensor arrays which are group invariant -- the so-called \emph{Markov invariants}. These include, for example, for the quartet tree case, the remarkable `squangles'; degree five polynomials which act as powerful quartet identifiers for the general Markov model, without the need for full parameter reconstruction \cite{sumner2009,holland2012}.

Our work on Markov invariants must be distinguished from related work on the similarly named \emph{phylogenetic invariants} \cite{lake1987,cavender1987,felsenstein1991,evans1993}.
Phylogenetic invariants are defined as those polynomials that vanish on a given phylogenetic tree (or subset of trees) under all (or nearly all) parameter settings of a given Markov model of sequence evolution.
As such, phylogenetic invariants form polynomial \emph{ideals} and hence can be analysed formally using algebraic geometry \cite{allman2004,sturmfels2004,draisma2008,casanellas2010}.
Beyond the theoretical significance of phylogenetic invariants (for example, they can be used to establish model identifiability \cite{allman2005}), the practical motivation behind the development of phylogenetic invariants lies in their vanishing (at least in expectation value) on particular trees.
Thus, when evaluated on an observed sequence alignment, phylogenetic invariants provide some information as to which evolutionary tree history the sequences are likely to have arisen from.

On the other hand, Markov invariants are defined as the one-dimensional polynomial \emph{representations} of the matrix group formed from the Markov matrices that act on the leaves of a phylogenetic tree. 
By definition, each Markov invariant spans a one-dimensional invariant subspace under changes of model parameter settings at the leaves of the tree.
Hence, Markov invariants provide useful statistical information that is invariant to the independent stochastic processes that have occurred since phylogenetically related taxa diverged from one another.
Phylogenetic invariants do not share this invariance property, and it is our contention that, at least comparatively, Markov invariants will provide particularly robust statistical information (particularly if we consider the setting of finite length sequence alignments where stochastic errors become important).

In a study of rodent phylogeny, a hitherto un-noticed interesting regime of DNA substitution parameters was pointed out by Yap and Pachter \cite{pachter2004identification}. 
They identified in their analysis, a special case of the \texttt{GTR} parameters, wherein the substitution matrix becomes invariant under Watson-Crick base conjugation (in consequence, the stationary base frequencies also satisfy $\pi_A\!=\!\pi_T$, $\pi_C\!=\!\pi_G$, consistent with Chargaff's rule). This model class was formally introduced as the `strand symmetric' model, and its defining ideals in the algebraic geometry approach considered in detail by \cite{casanellas2005strand,casanellas2010}.

This article focusses exclusively on a representation theoretic approach to the strand symmetric model and the derivation of Markov invariants for this model.
This is achieved by exploring a formal algebraic analysis of the Lie algebra associated with the model.
In \S\ref{sec:strandsymm}, we provide an abstract decomposition of this Lie algebra in terms of the Lie algebras of classical groups \cite{weyl1950}, and identify the particular representation provided by the $4\times 4$ rate matrices making up strand symmetric model.
In \S\ref{sec:markovinvs}, we couple our previous work characterising Markov invariants for the general Markov model \cite{sumner2008}, and our analysis of the underlying Lie algebra in \S\ref{sec:strandsymm}, to provide a complete classification and enumeration of binary and cubic Markov invariants for the strand symmetric model. 
The most technical aspects of the classification and enumeration of Markov invariants -- relying heavily on specialised manipulations of symmetric function characters (plethysm and skew operations) -- are relegated to the appendix \S\ref{sec:Appendix}; the casual reader should be able to follow the explicit construction of the invariants, without the need to fully understand the combinatorial derivations underlying our enumerations.
In \S\ref{sec:evaluate}, we examine the evaluation of quadratic Markov invariants on for a two-leaf phylogenetic tree. 
In this case there are four quadratic Markov invariants, which we show provide the means for estimating \emph{two} pairwise phylogenetic distances: constructed from the total of substitution rates \emph{within}, and \emph{across}, Watson-Crick conjugate base pairs, respectively. 
In the discussion \S\ref{sec:discussion}, we give concluding remarks and possibilities for future work including a comparison of the relative statistical power of phylogenetic and Markov invariants to accurately recover evolutionary trees. 


\section*{Acknowledgement}

Part of this work was completed by PDJ under an Australian senior Fulbright scholarship (Department of Statistics, University of California Berkeley, and Department of Physics, University of Texas at Austin) and hosts and colleagues at these institutions are thanked for their support. 
JGS was partially supported by Australian Research Council grants DP0877447, FT100100031, and DE130100423.
\section{The strand symmetric rate model and its Lie algebra structure}
\mbox{}
\label{sec:strandsymm}
The central construct in the standard theoretical approach to phylogenetic branching is an assumed substitution matrix parametrizing the probabilities for transitions between different states of a random variable which encodes the stochastic nature of biological molecular sequences (bases, for nucleic acids, or amino acids, for proteins). Concentrating on DNA, we have for example a 2 state system $\{ \texttt{R,Y}\}$ (purines and pyrimidines), or a 4 state system with state space $\{ \texttt{A,C,G,T}\}$. 

Consider firstly the two state case. The general 
Markov model in this case has substitution matrix 
\[
M = \left( \begin{array} {cc} 
m_{\texttt{RR}} & m_{\texttt{RY}}  \\
m_{\texttt{YR}} & m_{\texttt{YY}} \end{array}\right). 
\]
Probability conservation constrains each row of $M$ to have unit sum, so that there are two independent parameters 
$m_{\texttt{RY}} \equiv a$, $m_{\texttt{YR}} \equiv b$, with $M$ in the form
\[
{M(a,b)} = \left( \begin{array} {cc} 
1-a & a  \\
b  & 1-b \end{array}\right). 
\]
Noting the closure property given by the matrix multiplication rule 
\[
M(a,b)M(a',b')= M(a(1-a'-b')+a',b(1-a'-b')+b'),
\]
 we therefore characterize the general two-state Markov model as the set of substitution matrices $M(a,b)$ with $0\leq a,b\leq 1$ (technically a matrix semigroup). 
In order to apply group-theoretic methods, we enlarge the set $M(a,b)$ by working over the complex field and removing any constraints other than $\det(M(a,b))=1-a-b\neq 0$, thereby defining a certain matrix subgroup of the general linear group of nonsingular $2\!\times\! 2$ matrices. 
In the usual way, this group possesses a Lie algebra, its tangent space at the identity defined via derivatives and generated in this case by
$R_1:=\left.(\partial\!/\!\partial a) M(a,b)\right|_{a=b=0}$, $R_2:=\left.(\partial\!/\!\partial b) M(a,b)\right|_{a=b=0}$, namely\footnote{For aesthetic purposes here and below, signed entries in matrices are written with overbars.}
\begin{align}
R_1 = \left( \begin{array} {cc} 
\overline{1} & 1  \\
0  & 0 \end{array}\right), \qquad
R_2 =  \left( \begin{array} {cc} 
0 & 0  \\
1  & \overline{1} \end{array}\right),\nonumber
\end{align}
with the only non-trivial commutator bracket given by ${[}R_1,R_2{]}:=R_1R_2-R_2R_1= -R_1+R_2$. 
It is a general fact that, for arbitrary complex combinations $Q = \alpha R_1 + \beta R_2 $ in the Lie algebra, the matrix exponential $\exp(Q)$ belongs to the corresponding matrix group. In order to recover the Markov substitution model however, the off-diagonal matrix elements of such $Q$ should be positive quantities interpretable as substitution rates for the respective state transitions. Adopting a uniform normalization to negative unit trace, we characterize the two state Markov rate model as the set of matrices $M = \exp(tQ)$, $t>0$, with $Q = \alpha R_1 + \beta R_2 $ and
$\alpha, \beta \ge 0, \alpha + \beta =1$.

The situation for the general Markov substitution and rate models for the 4 state system, with state space $\{ \texttt{A,C,G,T}\}$, is similar. Allowing for the row sum constraint, the Markov matrix
\[
M = \left( \begin{array} {cccc} 
m_{\texttt{AA}} & m_{\texttt{AC}} & m_{\texttt{AG}} & m_{\texttt{AT}} \\
m_{\texttt{CA}} & m_{\texttt{CC}} & m_{\texttt{CG}} & m_{\texttt{CT}} \\
m_{\texttt{GA}} & m_{\texttt{GC}} & m_{\texttt{GG}} & m_{\texttt{GT}} \\
m_{\texttt{TA}} & m_{\texttt{TC}} & m_{\texttt{TG}} & m_{\texttt{TT}} 
\end{array} \right)
\]
has 12 free parameters, and the corresponding matrix group has Lie algebra spanned by $6+6=12$ standard generators analogous to $R_1$, $R_2$ above (two sets of six with positive unit entries above and below the diagonal, respectively, each with corresponding diagonal $-1$'s).
The general Markov rate model consists therefore of convex combinations of elements of the Lie algebra in the above basis (with nonnegative real coefficients), thus having negative unit trace.
Our interest here is in restricted model classes having the closure property, and the affiliated matrix subgroups of the general Markov model. The rate model for such a restricted class is then the intersection of the Lie subalgebra in question, with the general Markov rate model as above. We refer to these rate matrices as the stochastic cone of the Lie algebra\footnote{Consult \cite{sumner2012c}; details of the general case are not required in the present work.}. 

Consider now the general time reversible (\texttt{GTR}) model, where the guiding assumption is that transition rates involving arbitrary states $i,j\in\{\texttt{A},\texttt{C},\texttt{G},\texttt{T}\}$, weighted by the (stationary) distribution of the starting state $\pi_k$, are independent of whether the transition is from $i$ to $j$, or $j$ to $i$, technically stated as 
\[
\pi_i Q_{ij} = \pi_j Q_{ji}.
\]
In practice, this is implemented by taking an arbitrary \emph{symmetric} matrix $S$, and forming the (off diagonal) parts of $Q$ as the product of $S$ with the diagonal matrix of the stationary distribution,
\[
Q = \left( \begin{array} {cccc} 
Q_{\texttt{AA}} & S_{\texttt{AC}}\pi_{\texttt{C}} & S_{\texttt{AG}}\pi_{\texttt{G}} & S_{\texttt{AT}}\pi_{\texttt{T}} \\
S_{\texttt{CA}}\pi_{\texttt{A}} & Q_{\texttt{CC}} & S_{\texttt{CG}}\pi_{\texttt{G}} & S_{\texttt{CT}}\pi_{\texttt{T}} \\
S_{\texttt{GA}}\pi_{\texttt{A}} & S_{\texttt{GC}}\pi_{\texttt{C}} & Q_{\texttt{GG}} & S_{\texttt{GT}}\pi_{\texttt{T}} \\
S_{\texttt{TA}}\pi_{\texttt{A}} & S_{\texttt{TC}}\pi_{\texttt{C}} & S_{\texttt{TG}}\pi_{\texttt{G}} & Q_{\texttt{TT}} 
\end{array} \right),
\]
with $S_{ij}=S_{ji}$ and the diagonal entries set to ensure probability conservation  (zero row sums for rate matrices), for example
$Q_{\texttt{AA}} =- S_{\texttt{AC}}\pi_{\texttt{C}} - S_{\texttt{AG}}\pi_{\texttt{G}} - S_{\texttt{AT}}\pi_{\texttt{T}}$. As required, the row vector of stationary probabilities $(\pi_{\texttt{A}},\pi_{\texttt{C}},\pi_{\texttt{G}},\pi_{\texttt{T}})$ is a left null eigenvector of $Q$.

A special case of the \texttt{GTR} model occurs when its transition rates are unchanged under Watson-Crick base pairing conjugation (i.e. $\texttt{A}\leftrightarrow \texttt{T}$, $\texttt{C}\leftrightarrow \texttt{G}$); for example $Q_{\texttt{CA}} = Q_{\texttt{GT}}$,  $Q_{\texttt{CT}} = Q_{\texttt{GA}}$, $Q_{\texttt{TA}} = Q_{\texttt{AT}}$, and so on. In the above parametrization, imposition of this constraint on self-conjugate pairs
such as $Q_{\texttt{TA}} = Q_{\texttt{AT}}$ enforces Chargaff's rule on the stationary distribution, $\pi_{\texttt{A}} = \pi_{\texttt{T}}$, and $\pi_{\texttt{C}}=\pi_{\texttt{G}}$, and the remaining conditions constrain $S$ also to fulfil the analogous conditions $S_{\texttt{AC}} = S_{\texttt{GT}}$,  $S_{\texttt{AG}} = S_{\texttt{CT}}$ etc. (for self-conjugate pairs, the relations 
$S_{\texttt{CG}}= S_{\texttt{GC}}$ and $S_{\texttt{AT}}= S_{\texttt{TA}}$ are already enforced by the symmetry of $S$). As mentioned, the \texttt{GTR} model class is not multiplicatively closed \cite{sumner2011}, and neither will this base pairing conjugation symmetric case be. Remarkably however, the strand symmetric model, defined to fulfil the base pairing conjugation symmetry condition alone, does have the closure property, as follows. 

A convenient parametrization of the strand symmetric model occurs by fixing an arbitrary minimal set of transition probabilities, and duplicating these entries in the conjugate matrix elements.
Thus we choose
 \[
M = \left( \begin{array} {cccc} 
m_{\texttt{AA}} & m_{\texttt{AC}} & m_{\texttt{AG}} & m_{\texttt{AT}} \\
m_{\texttt{CA}} & m_{\texttt{CC}} & m_{\texttt{CG}} & m_{\texttt{CT}} \\
m_{\texttt{GA}} & m_{\texttt{GC}} & m_{\texttt{GG}} & m_{\texttt{GT}} \\
m_{\texttt{TA}} & m_{\texttt{TC}} & m_{\texttt{TG}} & m_{\texttt{TT}} 
\end{array} \right)
\equiv  \left(\begin{array}{cccc}
a & b & c & d \\
e & f & g & h \\
h & g & f & e \\
d & c & b & a \end{array}\right)
\]
where $a \equiv 1-b-c-d$, $f \equiv 1-e-g-h$. 
That closure indeed holds, follows trivially 
by verifying that the matrix product $MM'$ of two such patterned matrices respects the base conjugation symmetry.

In terms of the model classes referred to in the introductory discussion, the strand symmetric model occurs as an `equivariant model' \cite{draisma2008}, which is are useful generalisation of the standard `group-based' models (\cite{semple2003}, chapter 8) and are multiplicatively closed. 
Other examples are the Kimura three parameter model with
$b=e$, $c=h$, $g=d$, the Kimura two parameter model with $b=e=g=d$, $c=h$, and the Jukes-Cantor (one parameter) model with
$b=c=h=e=g=d$. 
In particular, the strand symmetric model is constructed as an equivariant model by including all substitution matrices $M$ invariant under simultaneous row and column permutations drawn from $\{\epsilon,(\texttt{A}\texttt{T})(\texttt{G}\texttt{C})\}$ (where $\epsilon$ is the identity or `do nothing' permutation), as is clear from the explicit form given above.  
As noted earlier, a broader approach to multiplicatively closed model classes, where the state space of the Markov chain is deemed to have some structure invariant under a fixed group of state permutations, has been presented in \cite{sumner2011,sumner2012c} under the banner of `Lie Markov' models.
In that work, a somewhat broader notion of model symmetry is utilized; where a model is deemed to have a certain permutation symmetry, not only if each individual substitution matrix is invariant under permutations drawn from the group (as in the equivariant case), but rather if each permutation produces a (possibly distinct) substitution matrix which is also included in the model. 
This notion of symmetry allows for permutations of individual parameters in the model, which, as is argued in \cite{sumner2011}, is consistent with the fact that the parameter labels play no intrinsic role, as parameters must be fitted to data using statistical inference.
In particular, in \cite{sumner2012c} a complete hierarchy consisting of 35 multiplicatively closed models  is derived\footnote{The exact number of models in the hierarchy depends somewhat on whether certain special cases are included in the count or not. The complete hierarchy, together with full details of matrix elements for each model, is provided online at \texttt{www.pagines.ma1.upc.edu/~jfernandez/LMNR.pdf}.}, which are additionally invariant under the permutations which fix the partitioning of nucleotides into purines and pyrimidines, i.e. $\texttt{A}\texttt{G}|\texttt{C}\texttt{T}:=\{\{\texttt{A},\texttt{G}\},\{\texttt{C},\texttt{T}\}\}$, so notationally $\texttt{A}\texttt{G}|\texttt{C}\texttt{T}\equiv \texttt{G}\texttt{A}|\texttt{C}\texttt{T} \equiv \texttt{T}\texttt{C}|\texttt{A}\texttt{G}\ldots $ etc.
In \cite{sumner2012c} it is also noted that an equivalent hierarchy exists for the partitioning that defines the Watson-Crick base pairing conjugation, i.e. $\texttt{A}\texttt{T}|\texttt{G}\texttt{C}$ (and yet another hierarchy for the partitioning $\texttt{A}\texttt{C}|\texttt{G}\texttt{T} $).
In particular, Model 6.6 \cite{sumner2012c} is identical to the strand symmetric model with the substitution $\texttt{G}\leftrightarrow \texttt{T} $ (or $\texttt{A}\leftrightarrow \texttt{C}$).
From this point of view, the strand symmetric model lives in a large hierarchy of Lie Markov models, equivalent to the hierarchy presented in \cite{sumner2012c}, where each model has symmetry consistent with Watson-Crick base pairing.

\begin{figure}
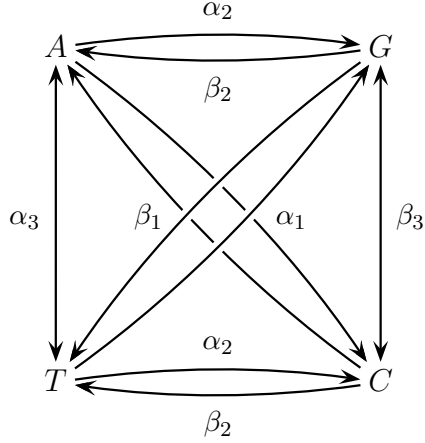

\centering
\vspace{1em}
$
\psmatrix[colsep=4cm,rowsep=4cm,linewidth=.03cm]
\text{\large{$A$}}&\text{\large{$G$}}\\
\text{\large{$T$}}&\text{\large{$C$}}
\ncarc[arcangle=8,border=2pt]{->}{1,2}{1,1}_{\beta_2}
\ncarc[arcangle=8]{->}{1,1}{1,2}^{\alpha_2}
\ncarc[arcangle=8,border=2pt]{->}{2,2}{2,1}_{\beta_2}
\ncarc[arcangle=8]{->}{2,1}{2,2}^{\alpha_2}
\ncline{<->}{1,1}{2,1}<{\alpha_3}
\ncline{<->}{1,2}{2,2}>{\beta_3}
\ncarc[arcangle=10,border=2pt]{->}{1,1}{2,2}>{\alpha_1}
\ncarc[arcangle=-10,border=2pt]{<-}{1,1}{2,2}<{\beta_1}
\ncarc[arcangle=10,border=2pt]{<-}{2,1}{1,2}
\ncarc[arcangle=-10,border=2pt]{->}{2,1}{1,2}
\endpsmatrix
$
\vspace{1em}
\caption{Graphical representation of the strand symmetric model.}
\label{fig:ssm}
\end{figure}

Let $\mathcal{S}$ be the vector space associated with the four nucleotide bases, with standard unit vectors 
\[e_\texttt{A}=\left(1,0,0,0\right),\quad e_\texttt{C}=\left(0,1,0,0\right),\quad e_\texttt{G}=\left(0,0,1,0\right),\quad e_\texttt{T}=\left(0,0,0,1\right),\]
so ${\mathcal S} := \langle e_\texttt{A}, e_\texttt{C}, e_\texttt{G}, e_\texttt{T} \rangle_{\mathbb{C}} \cong 
{\mathbb C}^4 $ and, for example, the state distribution $\pi_i$ is given by the vector $\pi=\pi_\texttt{A}e_\texttt{A}+\pi_\texttt{C}e_\texttt{C}+\pi_\texttt{G}e_\texttt{G}+\pi_\texttt{T}e_\texttt{T}$.
Following the analysis in the two state case, we consider the matrix Lie group affiliated to the strand symmetric model.
In the usual way of extracting the Lie algebra as the tangent space at the identity, we find,  in direct correspondence with variations in the independent parameters  $b,c,d,e,g,h$,  the following six generators:
\begin{align}
S_1 = & \, \left(\begin{array}{cccc}
\bar{1} & 1 & 0 & 0 \\
0 & 0 & 0 & 0 \\
0 & 0 & 0 & 0 \\
0 & 0 & 1 & \bar{1} \end{array}\right), \qquad 
S_2 = & \, \left(\begin{array}{cccc}
\bar{1} & 0 & 1 & 0 \\
0 & 0 & 0 & 0 \\
0 & 0 & 0 & 0 \\
0 & 1 & 0 & \bar{1} \end{array}\right), \qquad 
S_3 = & \, \left(\begin{array}{cccc}
\bar{1} & 0 & 0 & 1 \\
0 & 0 & 0 & 0 \\
0 & 0 & 0 & 0 \\
1 & 0 & 0 & \bar{1} \end{array}\right), \nonumber \\
T_1 = & \, \left(\begin{array}{cccc}
0 & 0 & 0 & 0 \\
1& \bar{1} & 0 & 0 \\
0 & 0 & \bar{1} & 1 \\
0 & 0 & 0 & 0 
\end{array}\right), \qquad 
T_2 = & \, \left(\begin{array}{cccc}
0 & 0 & 0 & 0 \\
0& \bar{1} & 0 & 1 \\
1 & 0 & \bar{1} & 0 \\
0 & 0 & 0 & 0 
\end{array}\right),
 \qquad 
 T_3 = & \, \left(\begin{array}{cccc}
0 & 0 & 0 & 0 \\
0& \bar{1} & 1 & 0 \\
0 & 1 & \bar{1} & 0 \\
0 & 0 & 0 & 0 
\end{array}\right).
 \nonumber 
&  
\end{align}
In this way, we can represent a rate matrix $Q$ as
\[
Q=\alpha_{1}S_1+\alpha_{2}S_2+\alpha_{3}S_3+\beta_{1}T_1+\beta_{2}T_2+\beta_{3}T_3,
\]
where $\alpha_1,\alpha_2,\alpha_3$ and $\beta_1,\beta_2,\beta_3$ are generic parameters.
Moreover, it is easily checked that 
the \emph{Ansatz} $\pi_{\texttt{A}} = \pi_{\texttt{T}} = p$, $\pi_{\texttt{C}}=\pi_{\texttt{G}} =q$  provides a left null eigenvector of the transition matrix $Q$, if $p= {(\beta_1+\beta_2)}/{2(\alpha_1+\alpha_2+\beta_1+\beta_2)}$, $q=  {(\alpha_1+\alpha_2)}/{2(\alpha_1+\alpha_2+\beta_1+\beta_2)}$
-- independently of $\alpha_3$ and $\beta_3$ -- which is therefore the unique stationary distribution.
A graphical representation of the model is given in Figure~\ref{fig:ssm}.

The full set of 15 commutation relations amongst these generators is
\begin{align}
{[}S_1,S_2{]}= & \, S_1-S_2, \qquad  {[}S_2,S_3{]}=  -S_1+S_2, \qquad {[}S_3,S_1{]}= -S_1+S_2, \nonumber \\
{[}T_1,T_2{]}=& \, T_1-T_2,  \qquad  {[}T_2,T_3{]}=  -T_1+T_2, \qquad  {[}T_3,T_1{]}=   \, -T_1+T_2,\nonumber \\
{[}S_1, T_1{]} =& \,  -S_1+T_1, \quad {[}S_1, T_2{]} =  -S_1+S_3 -T_3+T_2, \quad  {[}S_1, T_3{]} =  -S_1+S_2,\nonumber \\
{[}S_2, T_1{]} =& \,  -S_2+S_3 +T_1-T_3, \quad {[}S_2, T_2{]} =  -S_2+T_2, \quad  {[}S_2, T_3{]} =  S_1-S_2,\nonumber \\
{[}S_3, T_1{]} =& \,  T_1-T_2, \quad {[}S_3, T_2{]} =   -T_1+T_2, \qquad  {[}S_3, T_3{]} =  0,\nonumber 
\end{align}
as can be checked by elementary matrix algebra. 
We denote the corresponding complex Lie algebra by 
$l_{\texttt{SSM}}:=\langle S_1,S_2,S_3,T_1,T_2,T_3\rangle_\mathbb{C}$.

The group of permutations $\{\epsilon,(\texttt{A}\texttt{T}),(\texttt{G}\texttt{C}),(\texttt{A}\texttt{T})(\texttt{G}\texttt{C}),(\texttt{A}\texttt{G})(\texttt{C}\texttt{T}),(\texttt{A}\texttt{C})(\texttt{G}\texttt{T}),(\texttt{A}\texttt{G}\texttt{T}\texttt{C}),(\texttt{A}\texttt{T}\texttt{G}\texttt{C})\}$
fix the Watson-Crick pairing $\texttt{A}\texttt{T}|\texttt{G}\texttt{C}$, and are generated, for example, by the permutation $(\texttt{A}\texttt{T})$, via $\texttt{T}\texttt{A}|\texttt{G}\texttt{C}\!\equiv\! \texttt{A}\texttt{T}|\texttt{G}\texttt{C}$, and the permutation $(\texttt{A}\texttt{G})(\texttt{T}\texttt{C})$, via $\texttt{G}\texttt{C}|\texttt{A}\texttt{T}\!\equiv\! \texttt{A}\texttt{T}|\texttt{G}\texttt{C}$.
In terms of the generators of the Lie algebra $l_{\texttt{SSM}}$, these permutations produce the label substitutions $1\leftrightarrow 2$ and $S\leftrightarrow T$, respectively.

We now proceed via Levi's theorem \cite{erdmann2006} to give the structure of $l_{\texttt{SSM}}$ as the direct sum of a semisimple and a solvable part using the following matrix notation. 
We denote the unique three-dimensional simple Lie algebra ($A_1\cong B_1 \cong C_1$ in Cartan's classification) as $sl_2$, and the one-dimensional (abelian) Lie algebra ($\cong {\mathbb C}$ as a vector space) as $gl_1$.
As generators of the so-called `defining' representation of $sl_2$, with corresponding module $\mathcal{U}\cong \mathbb{C}^2$, we take:
\[
K_+=  \left( \begin{array} {cc} 
0 & 1  \\
0 & 0 \end{array}\right), \qquad
K_- =  \left( \begin{array} {cc} 
0 & 0  \\
1  & 0 \end{array}\right), \qquad
K_0=  \textstyle{\frac 12}\left( \begin{array} {cc} 
1 & 0  \\
0  & \overline{1} \end{array}\right),
\]
with commutation relations ${[}{K}_0,{K}_\pm{]} = \pm {K}_\pm$, ${[}{K}_+,{K}_-{]} = 2{K}_0$.
The Lie algebra affiliated to the two-dimensional general Markov model, with generators $R_1$ and $R_2$ described above, is isomorphic to the unique nonabelian two-dimensional Lie algebra \cite{erdmann2006}, consisting of the semidirect sum of a one-dimensional abelian algebra with a one-dimensional factor (often referred to as the `shift algebra'), generated by $X$ and $Y$ with non-zero commutation relation ${[}X,Y{]}=Y$. 
We denote this Lie algebra by $l_2$ and consider the representation\footnote{See \cite{sumner2010} for an algebraic investigation of the role of $l_2$ in phylogenetic tree and network models.} obtained by taking the module $\mathcal{V}\cong \mathbb{C}^2$ and generators $X \!=\! \textstyle{\frac 12}\left(R_1+R_2\right)$ and $Y\!=\textstyle{\frac 12}\left(\!R_2-R_1\right)$, i.e.
\begin{align}
X=\textstyle{\frac 12}\left(\begin{matrix}\overline{1} & 1 \\ 1 & \overline{1}\end{matrix}\right),\quad 
Y=\textstyle{\frac 12}\left(\begin{matrix}1 & \overline{1} \\ 1 & \overline{1}\end{matrix}\right).\nonumber
\end{align}

Note any module ${\mathcal W}$ of $sl_2\oplus l_2$ extends to a module ${\mathcal W}_r := {\mathcal W}\otimes {\mathcal R} \cong {\mathcal W}$ of $sl_2 \oplus l_2 \oplus gl_1$, where ${\mathcal R} \cong {\mathbb C} \cong \langle v \rangle_{\mathbb{C}}$ and $v$ is an eigenvector of a generator $R$ of $gl_1$ with eigenvalue $r$.
Below we will also have recourse to refer to the ``trivial'' representations of $sl_2$ and $gl_1$, with modules $\mathcal{U}_0\cong \mathbb{C}$ and $\mathcal{R}_0 \cong \mathbb{C}$ respectively, obtained by mapping all generators to 0.
We also require an additional representation of $l_2$ with corresponding module $\mathcal{V}'=\langle v' \rangle_\mathbb{C}\cong \mathbb{C}$, where $v'$ is an eigenvector for $X$ and is annihilated by $Y$.
\\

\noindent 
\textbf{Lemma 1}: \textbf{Decomposition of the Lie algebra of the strand symmetric model} \\
The Lie algebra $l_{\texttt{SSM}}$ generated by $S_1,S_2,S_3, T_1,T_1,T_2$ is isomorphic to the direct sum $sl_2\oplus gl_1 \oplus l_2$ of the simple three-dimensional Lie algebra $sl_2$, a one-dimensional Lie algebra $gl_1$, and the two-dimensional shift algebra $l_2$.
\\

\noindent
\textbf{Proof}: Define the new set of generators, 
\begin{align}
\widehat{K}_0 = & \,  \textstyle{\frac 14}(-S_3+T_3), \quad \widehat{K}_+  = \textstyle{\frac 12}(S_1-S_2), \quad \widehat{K}_-  = \textstyle{\frac 12}(T_1-T_2);
\nonumber \\
\widehat{R} = & \,  \textstyle{\frac 12}(S_3+T_3); \nonumber \\
\qquad \widehat{X} = & \, \textstyle{\frac 14}( S_1\!+\!S_2\!+\!T_1\!+\!T_2), \quad 
\quad \widehat{Y} =   \textstyle{\frac 14}(-S_1\!-\!S_2\!+\!S_3\!+\!T_1\!-\!T_3\!+\!T_2). \nonumber
\end{align}

By direct computation, 
\begin{align}
\widehat{K}_0 = & \, \textstyle{\frac 14}\left(\begin{array}{cccc}
1 & 0 & 0 & \bar{1} \\
0 & \bar{1} & 1 & 0 \\
0 & 1 & \bar{1} & 0 \\
\bar{1} & 0 & 0 & 1 \end{array}\right),
 \quad \widehat{K}_+ = 
\textstyle{\frac 12}\left(\begin{array}{cccc}
0 & 1 & \bar{1} & 0 \\
0 & 0 & 0 & 0 \\
0 & 0 & 0 & 0 \\
0 & \bar{1} & 1 & 0 \end{array}\right), \quad \widehat{K}_-=\textstyle{\frac 12}\left(\begin{array}{cccc}
0 & 0 & 0 & 0 \\
1 & 0 & 0 & \bar{1} \\
\bar{1} & 0 & 0 & 1\\
0 & 0 & 0 & 0  \end{array}\right); \nonumber \\
\widehat{R}= & \, \textstyle{\frac 12}\left(\begin{array}{cccc}
\bar{1} & 0 & 0 & 1 \\
0 & \bar{1} & 1 & 0 \\
0 & 1 & \bar{1} & 0 \\
1 & 0 & 0 & \bar{1} \end{array}\right); 
\quad \widehat{X} = \textstyle{\frac 14}\left(\begin{array}{cccc}
\bar{2} & 1 & 1 & 0 \\
1 & \bar{2} & 0 & 1 \\
1 & 0 & \bar{2} & 1 \\
0 & 1 & 1 & \bar{2} \end{array}\right), \quad
\widehat{Y} = \textstyle{\frac 14}\left(\begin{array}{cccc}
1 & \bar{1} & \bar{1} & 1 \\
1 & \bar{1} & \bar{1} & 1 \\
1 & \bar{1} & \bar{1} & 1 \\
1 & \bar{1} & \bar{1} & 1 \end{array}\right), \nonumber
\end{align}
we find the only non-zero commutation relations are ${[}\widehat{K}_0,\widehat{K}_\pm{]} = \pm \widehat{K}_\pm$, ${[}\widehat{K}_+,\widehat{K}_-{]} = 2\widehat{K}_0$, and ${[}\widehat{X},\widehat{Y}{]} = \widehat{Y}$, as required.\\
\mbox{} \hfill $\Box$
\\

\noindent 
\textbf{Lemma 2}: \textbf{Decomposition of the state space $\mathcal{S}$ of strand symmetric model} \\
As a module of $l_{\texttt{SSM}}\cong sl_2\oplus gl_1 \oplus l_2$, the four-dimensional state space $\mathcal{S}$ decomposes as the direct sum of two two-dimensional components $\mathcal{S}=U\oplus V$ where 
\begin{align}
U &\cong \left({\mathcal U}\otimes \mathcal{R}\otimes \mathcal{V}' \right),\nonumber\\ 
V &\cong\left(\mathcal{U}_0\otimes \mathcal{R}_0 \otimes \mathcal{V}\right),\nonumber
\end{align}
and 
\begin{enumerate}
\item $\mathcal{U}\cong \mathbb{C}^2$ and $\mathcal{U}_0\cong \mathbb{C}^1$ are the $sl_2$ modules described above,
\item $\mathcal{R}\cong \mathbb{C}$ and $\mathcal{R}_0\cong \mathbb{C}$ are the $gl_1$ modules described above,
\item $\mathcal{V}\cong \mathbb{C}^2$ and $\mathcal{V}'\cong \mathbb{C}$ are the $l_2$ modules described above.\\
\end{enumerate}

\noindent
\textbf{Proof}: As an alternative ordered basis for $\mathcal{S}$, take $\{u_0,u_1,v_{0'}, v_{1'}\}$\footnote{Here and below we will mark the indices of vectors (and/or tensor components) in $V$ by ${}'$.} where
\begin{align}
u_0 = \left(1,0,0,\overline{1}\right), 
\quad u_1 = \left(0,1,\overline{1},0\right), 
\quad v_{0'}   = \left(1,0,0,1\right),
\quad v_{1'}   = \left(0,1,1,0\right).
\nonumber 
\end{align}
\normalsize
\mbox{} \\
%

By direct computation, taking $\{u_0,u_1,v_{{0'}},v_{1'}\}$ as an ordered basis, we have the block forms (where boldface $\textbf{1}$ denotes the $2\times 2$ identity matrix):

\begin{equation}
\begin{tabular}{ccc}
$
\widehat{K}_0
=
\left(
\begin{matrix}
K_0 & 0 \\
0 & 0 \\
\end{matrix}
\right),
$ 

&

$
\widehat{K}_+=
\left(
\begin{matrix}
K_+ & 0 \\
0 & 0 \\
\end{matrix}
\right),
$ 

&

$
\widehat{K}_-=
\left(
\begin{matrix}
K_- & 0 \\
0 & 0 \\
\end{matrix}
\right),
$

\vspace{.5em}
\\

$
\widehat{R}=
\begin{pmatrix}
-\textbf{1} & 0 \\
0 & 0
\end{pmatrix},
$

&

$
\widehat{X}=
\begin{pmatrix}
-\textstyle{\frac{1}{2}}\textbf{1} & 0 \\
0 & X
\end{pmatrix},
$

&
 
$
\widehat{Y}=
\begin{pmatrix}
0 & 0 \\
0 & Y
\end{pmatrix}.
$

\end{tabular}
\nonumber
\end{equation}

Inspection of the blocks completes the proof.
\\
\mbox{}\hfill $\Box$

In the applications below we will refer to $\{u_0,u_1,v_{0'},v_{1'}\}$ as the `split' basis of $\mathcal{S}$.

\section{Application to Markov invariants}
\label{sec:markovinvs}
Our aim thus far has been to present the strand symmetric model \cite{pachter2004identification,casanellas2005strand} from the point of view of the underlying continuous Lie group.
In Lemma~1 we gave a classical decomposition of the Lie algebra associated with the strand symmetric model, and established the remarkable block diagonal form presented in Lemma~2.
We now turn to applications of these results.

In the analysis of \cite{casanellas2005strand}, the self-similar structure of the substitution matrices was exploited to 
formulate the strand symmetric model as a generalization from group-based models  to matrix valued group based models. 
This allows known Fourier/Hadamard inversion techniques to be pursued,
and in different situations, the ideal structure of the appropriate algebraic varieties can be described (including 
generalizations of the linear invariants, well-known from the vanishing coefficients in the Fourier basis occurring in the standard
group-based models).

Our approach with Lie group methods provides complementary insights. 
From the point of view of distance measures for phylogenetic reconstruction, any model can be subjected to tools such as the \texttt{LogDet} \cite{barry1987,lake1994,lockhart1994}.
The \texttt{LogDet} arises as a particular example of the more general concept of \emph{Markov invariants} \cite{sumner2008}, which are polynomials providing one-dimensional representations of the Lie group underlying a given phylogenetic model.
However, the great numerical appeal of the linear inversions, that the Hadamard conjugation provides for the Kimura three parameter model \cite{hendy1994}, is the availability of phylogenetic information via nothing more than a change of basis.
This should be compared to the polynomial calculations (as required by Markov invariants when the underlying Lie group arises from the general Markov model of sequence evolution), which are inherently more susceptible to stochastic error. 
In the case of Markov models with additional special symmetries, such as the strand symmetric model, it is of significant benefit that lower degree Markov invariants provide equivalent information to the \texttt{LogDet} (which, for a state space of size four such as DNA, is a degree 4 polynomial). 
As the  matrix Lie group underlying the strand symmetric model is nonabelian (as exhibited by non-zero commutation relations in $l_{\texttt{SSM}}$), a complete set of linear invariants is not available (and hence no linear inversion technique analagous to the Hadamard conjugation is applicable); however, it turns out that a hierarchy of quadratic Markov invariants can be deployed for any number of leaves. 

The following is derived in the appendix, \S\ref{sec:Appendix}, which relies on our previously established rules for working out the appropriate representations \cite{Jarvis2012adinvth} using calculations in the ring of symmetric functions. Here we quote the main result:\\

\noindent 
\textbf{Theorem 1}: \textbf{Count of quadratic Markov invariants for the strand symmetric model} \\
For $L$ leaves there are precisely $\textstyle{\frac 12}(3^L + (-1)^L)$ linearly independent quadratic Markov invariants for the strand symmetric model, namely $1,5,13,41,\cdots$ for $L=1,2,3,4,\cdots$ respectively\footnote{Integer sequence A046717 (see \texttt{http://oeis.org/}).}.
\\


\noindent
\textbf{Proof}: See \S\ref{sec:Appendix} below.
\mbox{}\hfill $\Box$
\\


In the $L=2$ leaf case the quadratic invariants are proxies for determinant functions, not of the full $4\! \times\! 4$ probability array, but for its $2 \! \times\!  2$ blocks in the split basis provided by the decomposition of the state space given in Lemma 2, and as such can provide differential information about the relative contributions of the rate parameters to the total edge lengths. 
In particular, these invariants provide a method for estimating the total sum of rates $\left(
\alpha_3+\beta_3\right)$ multiplied by time elapsed \emph{within} the Watson-Crick conjugate pairs, and the total sum of rates $\left(\alpha_1+\beta_1+\alpha_2+\beta_2\right)$ multiplied by time elapsed \emph{across} the Watson-Crick conjugate pairs (refer to Figure~\ref{fig:ssm} for illustration).
This is realized in the explicit constructions below in \S\ref{sec:evaluate} and should be compared to application of the \texttt{LogDet} \cite{barry1987,lake1994,lockhart1994}, which, when considered as a distance measure for the strand symmetric model, conjoins these two quantities into a total sum. 

In general, the $\textstyle{\frac 12}(3^L + (-1)^L)$ quadratic Markov invariants can be constructed as follows.
We work in the split basis $\{u_0,u_1,v_{0'},v_{1'}\}$ and define the anti-symmetric tensors 
\begin{align}
\epsilon_{ij}:=\left\{
\begin{matrix}
\phantom{-}1\text{ if }i=0,j=1,\\
-1\text{ if }i=1,j=0,\\
\phantom{-}0,\hspace{1.5em}\text{ otherwise};
\end{matrix}
\right.
\qquad
\overline{\epsilon}_{ij}:=\left\{
\begin{matrix}
\phantom{-}1\text{ if }i=0',j=1',\\
-1\text{ if }i=1',j=0',\\
\phantom{-}0,\hspace{2em}\text{ otherwise}.
\end{matrix}
\right.\nonumber
\end{align}
Given the block form $M=m\oplus \overline{m}$ of a strand symmetric model substitution matrix in the split basis, we have
\begin{align}
\det(m)&=\sum_{i_1,i_2,j_1,j_2=0,1}M_{i_1i_2}M_{j_1j_2}\epsilon_{i_1j_1}\epsilon_{i_2j_2},\nonumber
\\\
\det(\overline{m})&=\sum_{i_1,i_2,j_1,j_2=0',1'}M_{i_1i_2}M_{j_1j_2}\overline{\epsilon}_{i_1j_1}\overline{\epsilon}_{i_2j_2}.\nonumber
\end{align}

Choose an integer $q\leq L$ and a sequence $(a_1,a_2,\ldots,a_q)$ with $a_j\in \{1,2\}$, and consider the quadratic function $f^{(a_1,a_2,\ldots a_q)}$ on $L$-way tensors $\psi_{i_1i_2i_3\ldots i_L}$ defined by:
\begin{align}
f^{(a_1,a_2,\ldots a_q)}(\psi):=\sum
\psi_{i_1i_2i_3\ldots i_q0'0'\ldots 0'}\psi_{j_1j_2j_3\ldots j_q0'0'\ldots 0'}\epsilon^{(a_1)}_{i_1j_1}\epsilon^{(a_2)}_{i_2j_2}\ldots \epsilon^{(a_q)}_{i_qj_q},\nonumber
\end{align}
where $\epsilon^{(1)}_{ij}\!\equiv\! \epsilon_{ij}$ and $\epsilon^{(2)}_{ij}\!\equiv\! \overline{\epsilon}_{ij}$ and the summation is over the values $0,1,0',1'$ for all indices appearing in the expression.
An explicit check shows that if $\psi\rightarrow \psi'=M_1\otimes M_2\otimes \ldots M_L\cdot \psi$, where each $M_i$ a strand symmetric model substitution matrix, we have
\begin{align}
f^{(a_1,a_2,\ldots, a_q)}(\psi')=\widehat{\zeta}f^{(a_1,a_2,\ldots, a_q)}(\psi),\nonumber
\end{align} 
with $\widehat{\zeta}:=\prod_{1\leq i\leq q} \det(m^{(a_i)}_{i})$ and $m^{(1)}_{i}\equiv m_{i}$ and $m^{(2)}_{i}\equiv \overline{m}_{i}$.
From the multiplicative property of the determinant, it follows that each such function provides a Markov invariant for the strand symmetric model. 
Allowing for analogous constructions utilizing different subsets of $q$ parts of the tensor $\psi_{i_1i_2\ldots i_L}$ (and marginalizing on the remaining $L\!-\!q$ parts), shows that we can construct 
\begin{align}
\sum_{q=0}^L\binom{L}{q}2^q,\nonumber
\end{align} 
Markov invariants in this way. 
However, it is easy to show, using the anti-symmetry of $\epsilon_{ij}$ and $\overline{\epsilon}_{ij}$,  that if $q$ is odd the construction gives the zero polynomial.
For example, for $L\!=\!3$, we have
\begin{align}
f^{(1,1,2)}(\psi)&=\psi_{000'}\psi_{111'}-\psi_{100'}\psi_{011'}+\psi_{110'}\psi_{001'}-\psi_{010'}\psi_{101'}-\psi_{001'}\psi_{110'}\nonumber\\
&\phantom{hello}+\psi_{101'}\psi_{010'}-\psi_{111'}\psi_{000'}+\psi_{011'}\psi_{100'}\nonumber\\
&= 0.\nonumber
\end{align}
Excluding the cases where $q$ is odd, we see that we have constructed
\begin{align}
\sum_{q=0}^L \binom{L}{2q}2^{2q}={\frac 13}(3^L+(-1)^L),\nonumber
\end{align} 
Markov invariants that clearly linear independent (since they have distinct weights $\widehat{\zeta}$) consistent with Theorem~1, as required.

The reader should note that the construction of the binary Markov invariants can easily be understood in intuitive terms by taking an  $L-$way tensor $\psi_{i_1i_2i_3\ldots i_L}$ and implementing two steps, as follows.
Firstly, we ``marginalize'' $(L-q)$ of the indices via,
\[
\psi_{i_1i_2i_3\ldots i_L}\rightarrow \psi_{i_1i_2i_3\ldots i_q0'0'\ldots 0'},
\]
where the $0'$ component simply expresses probability conservation in the underlying Markov chain. 
Secondly, we exploit the block form $M=m\oplus \overline{m}$ of the strand symmetric model by considering $m,\overline{m}\in GL(2)$ and ``saturate'' indices with the $GL(2)$ invariant tensors $\epsilon_{ij}$ and $\overline{\epsilon}_{ij}$.
As such, beyond the initial marginalization step, there is no direct exploitation of the more fined-grained observation that $\overline{m}$ actually belongs to a proper matrix-subgroup of $GL(2)$.

However, in the higher degree Markov invariants, things become combinatorially more interesting, as we now illustrate specifically for the cubic case.\newline

\noindent 
\textbf{Theorem 2}: \textbf{Count of cubic Markov invariants for the strand symmetric model} \\
For $L$ leaves there are precisely $6^{L-1}$ linearly independent cubic Markov invariants for the strand symmetric model, namely $1,6,36,216,\ldots$ for $L=1,2,3,4,\ldots$ respectively.
\\


\noindent
\textbf{Proof}:\\
See \S~\ref{sec:Appendix} below.
\mbox{}\hfill $\Box$
\\

As alluded to above, the explicit construction of the cubic invariants is not as straightforward as the the quadratic case. 
To illustrate, we give the complete list of 36 cubic linear independent Markov invariants for $L=3$.

Consider the three-way tensors $\psi_{i_1i_2i_3}$ with $\psi\mapsto M_1\otimes M_2\otimes M_3\cdot \psi$ where each $M_i=m_{i}\oplus \overline{m}_{i}$ belongs to the strand symmetric model.
We set $w_i:=\det(m_{i})$ and $\lambda_i:=\det(\overline{m}_{i})$.
The enumeration given in \S\ref{sec:Appendix} shows that the individual counts for various weights $\widehat{\zeta}$ of the 36 Markov invariants are given by
\begin{align}
1;\lambda_i\lambda_j;\lambda_1\lambda_2\lambda_3;\lambda_iw_j;2\lambda_i\lambda_jw_k;2w_iw_j;3\lambda_iw_jw_k;4w_1w_2w_3,\nonumber
\end{align}
for all choices $\{i,j,k\}=\{1,2,3\}$ and multiplicities have been included as a multiplicative factor, e.g. there is \emph{one} Markov invariant with weight $\lambda_1\lambda_2$ and there are \emph{three} Markov invariants with weight $\lambda_1w_2w_3$.

Using constructions inspired by the quadratic case above, we used Mathematica \cite{mathematica} to explicitly find 36 linearly independent Markov invariants, with the following results:

\begin{enumerate}
\item The \textbf{single} invariant with trivial weight $1$ is simply given by the cube of the probability sum: $\psi_{0'0'0'}^3$.

\item For the three quadratic weights of the form $\lambda_i\lambda_j$, the expression $\psi_{0'0'0'}\times \sum \psi_{i_1i_20'}\psi_{j_1j_20'}\overline{\epsilon}_{i_1j_1}\overline{\epsilon}_{i_2j_2}$ plus the obvious two permutations across the tensor indices provides the required \textbf{three} invariants.

\item For the quadratic weight $\lambda_1w_2$, the expression $\psi_{0'0'0'}\times \sum \psi_{i_1i_20'}\psi_{j_1j_20'}\overline{\epsilon}_{i_1j_1}\epsilon_{i_2j_2}$ provides the required invariant. 
Since there are six distinct  quadratic weights of the form $\lambda_iw_j$, this gives the required total of \textbf{six} invariants. 

\item For the quadratic weight $w_1w_2$, the expressions $\psi_{0'0'0'}\times \sum \psi_{0' i_2i_3}\psi_{0' j_2j_3}\epsilon_{i_2j_2}\epsilon_{i_3j_3}$ and  $\sum \psi_{0' 0' i_3}\psi_{0' j_20'}\psi_{0' k_2k_3}\epsilon_{j_2k_2}\epsilon_{i_3k_3}$ give two linearly independent invariants. 
Since there are three distinct  quadratic weights of the form $w_iw_j$, this gives the required total of \textbf{six} invariants. 

\item For the cubic weight $\lambda_1\lambda_2w_3$, the expression $\psi_{i_1i_2i_3}\psi_{j_1j_2 0'}\psi_{0' 0'k_3}\overline{\epsilon}_{i_1j_1}\overline{\epsilon}_{i_2j_2}\epsilon_{i_3k_3}$ plus two permutations provides only two linearly independent invariants.
Since there are three distinct cubic weights of the form $\lambda_i\lambda_jw_k$, this gives the required total of \textbf{six} invariants.

\item For the cubic weight $\lambda_1w_2w_3$, the expressions $\sum\psi_{i_1i_2i_3}\psi_{j_1j_20'}\psi_{0'0'k_3}\bar{\epsilon}_{i_1j_1}\epsilon_{i_2j_2}\epsilon_{i_3k_3}$,\newline $\sum\psi_{i_1i_2i_3}\psi_{j_10'j_3}\psi_{0'k_20'}\bar{\epsilon}_{i_1j_1}\epsilon_{i_2k_2}\epsilon_{i_3j_3}$ and $\sum\psi_{i_1i_2i_3}\psi_{0'j_2j_3}\psi_{k_10'0'}\bar{\epsilon}_{i_1k_1}\epsilon_{i_2j_2}\epsilon_{i_3j_3}$ provide three linearly independent invariants.
Since there are three distinct cubic weights of the form $\lambda_iw_jw_k$, this gives the required total of \textbf{nine} invariants.

\item For the cubic weight $\lambda_1\lambda_2\lambda_3$, the expression $\sum\psi_{i_1i_2i_3}\psi_{j_1j_20'}\psi_{0'0'k_3}\bar{\epsilon}_{i_1j_1}\bar{\epsilon}_{i_2j_2}\bar{\epsilon}_{i_3k_3}$ plus permutations across tensor indices provides only \textbf{one} linearly independent invariant, as required.

\item For the cubic weight $w_1w_2w_3$, the invariant $\sum\psi_{i_1i_2i_3}\psi_{j_1j_20'}\psi_{0'0'k_3}{\epsilon}_{i_1j_1}{\epsilon}_{i_2j_2}{\epsilon}_{i_3j_3}$ plus two permutations across tensor indices, plus the expression $\sum\psi_{0'i_2i_3}\psi_{j_10'j_3}\psi_{k_1k_20'}{\epsilon}_{j_1k_1}{\epsilon}_{i_2k_2}{\epsilon}_{i_3j_3}$ provide the required \textbf{four} invariants.

\end{enumerate}
 
The enumeration of the 36 Markov invariants is summarised in Table~\ref{tab:cubiclist}.
The reader should note that the first invariant is simply the trivial invariant cubed and the next three lines of invariants are all of the form (trival)$\times$(quadratic).
The remaining invariants are non-factorizable, and illustrate how the probability conservation invariance for different parts of the tensor is shared across different terms in the cubic product. 

As an example, consider the single invariant with weight $\lambda_1\lambda_2\lambda_3$:
\begin{align}
\sum\psi_{i_1i_2i_3}\psi_{j_10'j_3}\psi_{0'k_20'}\bar{\epsilon}_{i_1j_1}\bar{\epsilon}_{i_2k_2}\bar{\epsilon}_{i_3j_3}=
-&2\psi_{0'0'1'}\psi_{1'0'0'}\psi_{0'1'0'}
-\psi_{0'0'0'}^2\psi_{1'1'1'}\nonumber\\
&+\psi_{0'0'0'}\left(\psi_{0'1'1'}\psi_{1'0'0'}+
\psi_{1'0'1'}\psi_{0'0'1'}+
\psi_{1'1'0'}\psi_{0'0'1'}\right)\nonumber.
\end{align}
This invariant is of particular interest as it is already known in the context of the two-state general Markov model as both (i) a Markov invariant the ``stangle'' \cite{sumner2008}, and (ii) a three-way covariance on triplet trees \cite{klaere2012}.
The fact that this Markov invariant arises again in the case of the strand symmetric is remarkable, but should be expected given the two-part block decomposition given in Lemma 2. 
 
\begin{table}
\label{tab:cubiclist}
\begin{tabular}{cc|cc|cc}
Example Invariant & Weight &  Perms & Lin. indep. &  Weight perms & Total \\
\hline
\phantom{W}$\psi_{0'0'0'}^3$\phantom{W} & 1 & 1 & 1 & 1 & 1 \\
 $\psi_{0'0'0'}\times\sum\psi_{i_1i_20'}\psi_{j_1j_20'}\bar{\epsilon}_{i_1j_1}\bar{\epsilon}_{i_2j_2}$ & $\lambda_1\lambda_2$ & 1 & 1 & 3 & 3\\
 $\psi_{0'0'0'}\times\sum\psi_{i_1i_20'}\psi_{j_1j_20'}\bar{\epsilon}_{i_1j_1}\epsilon_{i_2j_2}$ &$\lambda_1w_2$ & 1 & 1 & 6 & 6  \\
 $\psi_{0'0'0'}\times\sum\psi_{0' i_2i_3}\psi_{0' j_2j_3}\epsilon_{i_2j_2}\epsilon_{i_3j_3}$ &$w_1w_2$ & 1 & 1 & 3 & 3 \\
$\sum\psi_{0'0'i_3}\psi_{0' j_20'}\psi_{0' k_2k_3}\epsilon_{j_2k_2}\epsilon_{i_3k_3}$ &$w_1w_2$ &  1 & 1 & 3 & 3 \\
 $\sum\psi_{i_1i_2i_3}\psi_{j_10'j_3}\psi_{0'k_20'}\bar{\epsilon}_{i_1j_1}\bar{\epsilon}_{i_2k_2}\epsilon_{i_3j_3}$ &$\lambda_1\lambda_2w_3$ & 3 & 2 & 3 & 6\\
 $\sum\psi_{i_1i_2i_3}\psi_{j_10'j_3}\psi_{0'k_20'}\epsilon_{i_1j_1}\epsilon_{i_2k_2}\bar{\epsilon}_{i_3j_3}$ &$\lambda_1w_2w_3$ & 3 & 3 & 3 & 9 \\
 $\sum\psi_{i_1i_2i_3}\psi_{j_10'j_3}\psi_{0'k_20'}{\epsilon}_{i_1j_1}{\epsilon}_{i_2k_2}{\epsilon}_{i_3j_3}$ &$w_1w_2w_3$ & 3 & 3 & 3 & 3 \\
 $\sum\psi_{0'i_2i_3}\psi_{j_10'j_3}\psi_{k_1k_20'}{\epsilon}_{j_1k_1}{\epsilon}_{i_2k_2}{\epsilon}_{i_3j_3}$ &$w_1w_2w_3$ & 1 & 1 & 1 & 1\\
$\sum\psi_{i_1i_2i_3}\psi_{j_10'j_3}\psi_{0'k_20'}\bar{\epsilon}_{i_1j_1}\bar{\epsilon}_{i_2k_2}\bar{\epsilon}_{i_3j_3}$ &$\lambda_1\lambda_2\lambda_3$ &  3 & 1 & 1 & 1 \\
\hline
 & & & & & \textbf{36}
\end{tabular} 
\caption{Cubic Markov invariants for 3-way tensors under the strand symmetric model.}
\end{table}

\section{Evaluation of the quadratic Markov invariants on phylogenetic trees}
\label{sec:evaluate}

The explicit evaluation of the explicit quadratic Markov invariants for $L=2$ proceeds as follows. 
We regard the probability pattern frequency array $\big( P_{ij}\big)_{i,j \in \{\texttt{A},\texttt{C},\texttt{G},\texttt{T}\}}$, as an element of ${\mathcal S} \otimes {\mathcal S}$, viz.
\[
P = \sum{}_{i,j} \, P_{ij} e_i \otimes e_j
\]
relative to the standard unit vectors $e_\texttt{A},e_\texttt{C},e_\texttt{G},e_\texttt{T}$ for ${\mathcal S} \cong {\mathbb C}^4$.
Via the transformation to the split basis $\{u_0,u_1,v_{0'}, v_{1'}\}$ we can write
\begin{align*}
\hskip-1.5ex\left(\small \begin{array}{cccc}
 P_{00} & P_{01} & P_{0 0'} & P_{01'} \\
P_{10} & P_{11} & P_{1 0'} & P_{11'} \\
P_{0'0} & P_{0'1} & P_{0'0'} & P_{0'1'} \\
P_{1'0} & P_{1'1} & P_{1' 0'} & P_{1'1'} 
\end{array} \normalsize \right)
\!=& \! \left( \begin{array}{cccc}
\hskip-1ex P_{\texttt{A}\texttt{A} \!-\! \texttt{A}\texttt{T}\!-\! \texttt{T}\texttt{A}\!+\!\texttt{T}\texttt{T}},
&\hskip-1ex P_{\texttt{A}\texttt{C}  \!-\!  \texttt{A}\texttt{G} \!-\!  \texttt{T}\texttt{C}  \!+\! \texttt{T}\texttt{G}},&
\hskip-1ex P_{\texttt{A}\texttt{A} \!+\! \texttt{A}\texttt{T}\!-\! \texttt{T}\texttt{A}\!-\!\texttt{T}\texttt{T}},&
\hskip-1exP_{\texttt{A}\texttt{C} \!+\!  \texttt{A}\texttt{G}\!-\!  \texttt{T}\texttt{C}  \!-\! \texttt{T}\texttt{G}} \\
\hskip-1ex P_{\texttt{C}\texttt{A}  \!-\!  \texttt{C}\texttt{T} \!-\!  \texttt{G}\texttt{A} \!+\! \texttt{G}\texttt{T}},
&\hskip-1ex P_{\texttt{C}\texttt{C}  \!-\!  \texttt{C}\texttt{G} \!-\!  \texttt{G}\texttt{C} \!+\! \texttt{G}\texttt{G}},&
\hskip-1ex P_{\texttt{C}\texttt{A}  \!+\!  \texttt{C}\texttt{T} \!-\!  \texttt{G}\texttt{A} \!-\! \texttt{G}\texttt{T}},&
\hskip-1exP_{\texttt{C}\texttt{C}  \!+\!  \texttt{C}\texttt{G} \!-\!  \texttt{G}\texttt{C} \!-\! \texttt{G}\texttt{G}}  \\
\hskip-1ex P_{\texttt{A}\texttt{A} \!-\! \texttt{A}\texttt{T}\!+\! \texttt{T}\texttt{A}\!-\!\texttt{T}\texttt{T}},
&\hskip-1ex P_{\texttt{A}\texttt{C}  \!-\!   \texttt{A}\texttt{G} \!+\!  \texttt{T}\texttt{C} \!-\! \texttt{T}\texttt{G}},&
\hskip-1ex P_{\texttt{A}\texttt{A} \!+\! \texttt{A}\texttt{T}\!+\! \texttt{T}\texttt{A}\!+\!\texttt{T}\texttt{T}},&
\hskip-1exP_{\texttt{A}\texttt{C}   \!+\!  \texttt{A}\texttt{G} \!+\!  \texttt{T}\texttt{C} \!+\! \texttt{T}\texttt{G}}
\\
\hskip-1ex P_{\texttt{C}\texttt{A}  \!-\!  \texttt{C}\texttt{T} \!+\!  \texttt{G}\texttt{A} \!-\! \texttt{G}\texttt{T}},
&\hskip-1ex P_{\texttt{C}\texttt{C}  \!-\!  \texttt{C}\texttt{G} \!+\!  \texttt{G}\texttt{C} \!-\! \texttt{G}\texttt{G}},&
\hskip-1ex P_{\texttt{C}\texttt{A}  \!+\!  \texttt{C}\texttt{T} \!+\!  \texttt{G}\texttt{A} \!+\! \texttt{G}\texttt{T}},&
\hskip-1exP_{\texttt{C}\texttt{C}  \!+\!  \texttt{C}\texttt{G} \!+\!  \texttt{G}\texttt{C} \!+\! \texttt{G}\texttt{G}} 
\end{array}   \right)\hskip-.5ex,
\nonumber
\end{align*}
where, for instance, $P_{\texttt{A}\texttt{A} \!+\! \texttt{A}\texttt{T}\!-\! \texttt{T}\texttt{A}\!-\!\texttt{T}\texttt{T}}:=P_{\texttt{A}\texttt{A}} \!+\!P_{ \texttt{A}\texttt{T}}\!-\! P_{\texttt{T}\texttt{A}}\!-\!P_{\texttt{T}\texttt{T}}$. 

Using the notation from above, the five quadratic Markov invariants for $L=2$ are given by (setting $q=0$) the trivial invariant $P^2_{0'0'}=\left(P_{\texttt{A}\texttt{A}}+P_{\texttt{A}\texttt{C}}+\ldots +P_{\texttt{T}\texttt{T}}\right)^2$ and (setting $q=2$) the four invariants  
\[
\begin{tabular}{ll}
$f^{(1,1)}(P)=P_{00}P_{11}-P_{01}P_{10}$,& $f^{(1,2)}(P)=P_{00'}P_{11'}-P_{01'}P_{10'}$,\\
$f^{(2,1)}(P)=P_{0'0}P_{1'1}-P_{0'1}P_{1'0}$, & $f^{(2,2)}(P)=P_{0'0'}P_{1'1'}-P_{0'1'}P_{1'0'}$,
\end{tabular}
\]
which can, of course, be recognised as determinants of the four $2\times 2$ blocks comprising $P$ in the split basis.
%

Letting $P\rightarrow P'=M_1\otimes M_2\cdot P$, we have 

\[
\begin{matrix}
f^{(1,1)}(P')=w_1w_2f^{(1,1)}(P),& f^{(1,2)}(P')=w_1\lambda_2f^{(1,2)}(P),\\
f^{(2,1)}(P')=\lambda_1w_2f^{(2,1)}(P),& f^{(2,2)}(P')=\lambda_1\lambda_2f^{(2,2)}(P).
\end{matrix}
\]

Let us evaluate these quadratic Markov invariants on a two-leaf phylogenetic tree. 
In the general case, parameterise the root distribution as $(p_\texttt{A}, p_\texttt{C}, p_\texttt{G}, p_\texttt{T}) := (p+r,q+s,q-s,p-r)$ with $p+q\!=\!1$. 
Immediately after speciation into two taxa we obtain the initial tensor $\tilde{P}$, which, in the standard basis has components: 
\[
\tilde{P}_{ij}=\left\{\begin{array}{l} p_i, \text{ if }i=j, \\ 0, \text{ otherwise;} \end{array}\right.
\]
for each $i,j=\texttt{A},\texttt{C},\texttt{G},\texttt{T}$, and hence in the split basis
\begin{align*}
\hskip-1.5ex\left(\small \begin{array}{cccc}
 \tilde{P}_{00} & \tilde{P}_{01} & \tilde{P}_{0 {0'}} & \tilde{P}_{0 {1'}} \\
\tilde{P}_{10} & \tilde{P}_{11} & \tilde{P}_{1 {0'}} & \tilde{P}_{1  {1'}} \\
\tilde{P}_{{0'}0} & \tilde{P}_{{0'}1} & \tilde{P}_{{0'} {0'}} & \tilde{P}_{{0'}  {1'}} \\
\tilde{P}_{{1'}0} & \tilde{P}_{{1'}1} & \tilde{P}_{{1'} 0'} & \tilde{P}_{{1'}  {1'}} 
\end{array} \normalsize \right)
=&  \frac{1}{2}\left(\small \begin{array}{cccc}
 p & 0 & r & 0 \\
0 & q & 0 & s \\ 
r & 0 & p & 0 \\
0 & s & 0 & q \\ 
\end{array} \normalsize \right).
\nonumber
\end{align*}

Extending to two-taxa tree with evolution of taxa 1 and described by strand symmetric transition matrices  $M_1$ and $M_2$, we obtain the phylogenetic tensor $P=M_1\otimes M_2\cdot \tilde{P}$. 
However, by a standard argument (the so-called `pulley-principle' \cite{felsenstein1981}), it is enough to evaluate the special case where after speciation one taxon remains fixed whilst the DNA of the other undergoes random substitutions.
Mathematically this allows us to set $M_1\equiv M$ and $M_2=I$. 
Explicit inspection then shows we have the values $f^{(1,1)}(P)= \frac{1}{4}w pq$, $f^{(1,2)}(P)=\frac{1}{4}w rs$, $f^{(2,1)}(P) = \frac{1}{4}\lambda rs$ and $ f^{(2,2)}(P) = \frac{1}{4}\lambda pq$.


Of course in the split basis, the strand symmetric Markov matrix $M$ is cast via Lemma~2 above  into the form of a direct sum of two $2\!\times\! 2$ blocks,
\[
{M} = \left( \begin{array} {cc} 
m & 0  \\
0  & \overline{m} \end{array}\right). 
\]
Correspondingly we have the weights $w=\det(m)$ and $\lambda=\det(\overline{m})$.
With Jacobi's formula $\det e^{Qt} = e^{tr(Q)t}$ in mind, inspection of the diagonal forms given in Lemma~3 shows that the only generators of the Lie algebra with non-zero trace are $\widehat{R}$ and $\widehat{X}$.
Considering the block form and evaluating matrix traces yields
\[
\det(m)=e^{-\left(2\sigma_1+\sigma_2\right)t},\qquad
\det(\overline{m})=e^{-\sigma_2t};
\]
where $\sigma_1:=\alpha_3+\beta_3$ and $\sigma_2:=\alpha_1+\alpha_2+\beta_2+\beta_3$ are the sum of rates \emph{within} and \emph{across} Watson-Crick pairs, respectively.
Thus on a two-taxa probability tensor $P$ arising under the strand symmetric model, we have the forms
\[
\begin{tabular}{lcl}
$f^{(1,1)}(P)=\frac{1}{4}e^{-(2\sigma_1-\sigma_2)t}pq$, && $f^{(1,2)}(P)=\frac{1}{4}e^{-(2\sigma_1+\sigma_2)}rs$, \\
$f^{(2,1)}(P)=\frac{1}{4}e^{-\sigma_2t}rs$, && $f^{(2,2)}(P)=\frac{1}{4}e^{-\sigma_2t}rs$.
\end{tabular}
\]
Thus, under the assumption of the strand symmetric model, and depending upon one's willingness to make assumptions about the root distribution parameters $p,q,r$ and $s$ (for example, assume a stationary distribution with $r\!=\!s\!=\!0$), it is possible to use the quadratic Markov invariants in a practical setting to obtain independent estimators of the overall within and across rates $\sigma_1$ and $\sigma_2$.

\section{Discussion}
\label{sec:discussion}

In this article we have explored the matrix group  properties of the strand symmetric model of DNA evolution from a representation theoretic point of view.
We gave a classical decomposition of the Lie algebra associated with the model  and further decomposed the representation of this Lie algebra occurring on DNA state space into irreducible modules.
We gave a full classification and enumeration of binary and cubic Markov invariants for this model. 
This work should be seen as distinct from, but complementary to, other results on the strand symmetric model taken from the point of view of ``algebraic statistics''.

Future work includes the examination of Markov invariants for the strand symmetric model on evolutionary trees with taxa greater than $L=3$.
Of particular interest, are the application of Markov invariants to the quartet case $L=4$.
The case of quartets is of special interest to applied phylogenetics, as this is smallest subset of taxa for which the evolutionary tree history is non-trivial (in the topological sense) relative to Markov models of sequence evolution.
Additionally, it is well known that it is enough to recover the evolutionary relations between all quartets of a set of taxa in order to be able to infer the the evolutionary tree of the full set (see \cite[Chap. 6]{semple2003} for the relevant discussion).
Similarly to the Markov invariants for the general Markov model, as studied in \cite{holland2012}, our initial results (unpublished) show that the Markov invariants for the strand symmetric model on quartets of taxa can be used effectively to infer phylogenetic trees.

We defer further speculation on these matters to future work.

\bibliography{masterC}
\bibliographystyle{plain}

\begin{appendix}
\section{Enumeration of Markov invariants for the strand symmetric model}
\label{sec:Appendix}
The following discussion adopts the notation and adapts the results of \cite{sumner2008,sumner2009,jarvis2010}, and especially \cite{Jarvis2012adinvth}.
The required background on symmetric function manipulations can be found in the classic text \cite{macdonald1979}.

In the language of representation theory, polynomials in $L$-way tensors $\psi_{i_1i_2\ldots i_L}$ are technically polynomial representations of the underlying matrix groups. 
For general matrix groups, our starting point is the representations of the general linear group $GL(n)$, or equivalently its Lie algebra $gl_n$, where the irreducible representations are labelled by (ordered) integer partitions $\lambda\vdash m$ with $\lambda=(\lambda_1,\lambda_2,\ldots,\lambda_r)$, $\lambda_1\geq \lambda_2 \geq  \ldots \geq \lambda_r\geq 0 $ and $\sum_{i}\lambda_i=m$. 
When a partition $\lambda$ labels directly (and equivalently) a particular $gl_n$ module, irreducible representation, or character, we adopt Littlewood's \cite{littlewood1940} notation for Schur functions, where the partition is enclosed by curly brackets: $\{\lambda\}$.

For a module ${\mathcal S}$ of a matrix group $G\leq GL(n)$, polynomials of degree $D$ in the components of ${\mathcal S}$ belong to the (in general reducible) module ${\mathcal S}\underline{\otimes} \{D\}$: the plethysm of ${\mathcal S}$ with the one-part partition $\lambda=(D)$. 
For an $L$-way phylogenetic pattern tensor, the module is the $L$-fold tensor product of the corresponding direct product group $G \!\times\! G \!\times\! \cdots \!\times\! G$ (one copy for each leaf on the phylogenetic tree). 
In this case the resolution of $\big( \otimes^L{\mathcal S}\big)\underline{\otimes}\{D\}$ requires calculation of generic plethysms ${\mathcal S}\underline{\otimes}\sigma$, where $\sigma \vdash D$. 
Further, the multiplicities $g^\lambda_{\mu \nu}$, with $\lambda,\mu,\nu\vdash D$, which resolve tensor products (inner multiplication `$\ast$') of irreducible modules in the symmetric group ${\mathfrak S}_D$, must also be computed.

The following is taken from \cite{Jarvis2012adinvth}:\\

\noindent 
\textbf{Lemma 4: General Enumeration of Markov invariants.} \\
To enumerate Markov invariants at degree $D$, carry out the following steps:\\
1. For each $\sigma \vdash D$, compute the number of one-dimensional representations $f_\sigma$ occurring
in the decomposition of ${\mathcal S}\underline{\otimes} \{\sigma\}$. \\
2. The number of Markov invariants at degree $D$ is then
\[
n_D = \sum_{\sigma_1,\sigma_2,\cdots, \sigma_L \vdash D} g^{(D)}_{\sigma_1\sigma_2\cdots \sigma_L}
f_{\sigma_1}f_{\sigma_2}\cdots f_{\sigma_L}
\]
where $g^{(D)}_{\sigma_1\sigma_2\cdots \sigma_L}$ is the inner product multiplicity for the occurrence of the module $(D)$ in the tensor product ${\sigma_1\otimes \sigma_2 \otimes \cdots \otimes \sigma_L}$ of modules of ${\mathfrak S}_D$.\\
\noindent
\mbox{}\hfill $\Box$

As a simple first case, we consider the enumeration of the quadratic, $D=2$, Markov invariants for the strand symmetric model.
\\

\noindent 
\textbf{Lemma 5}: \textbf{Calculation of $f_\sigma$ for $\mathcal{S}$ at degree $D=2$.} \\
At degree $D=2$ there are two partitions $\sigma\vdash D$ given by $(2)$ and $(1^2)$.
Symmetric function manipulations establish that $f_{(1^2)}=2$ and $f_{(2)}=1$.
\\

\noindent
\textbf{Proof}:
We appeal to the left-distributive law for plethysms \cite{littlewood1940}:
\begin{equation}
(A+B)\underline{\otimes} C=\sum _{\mu \subset C} \left(A\underline{\otimes} \, C/\mu\right)\otimes \left(B\underline{\otimes} \mu\right),\nonumber
\end{equation}
where $A,B,C$ are $gl_n$ characters and the summation is over all $\mu$ where the skew character $C/\mu$ is defined.
Referring to Lemma~2 and ignoring the modules $\mathcal{R},\mathcal{R}_0,\mathcal{V}'$ and $\mathcal{U}_0$, which being one-dimensional do not influence our calculation,  we take $A\equiv \mathcal{U}$ as a $sl_2< gl_2$ module and $B\equiv \mathcal{V}$ as a $l_2< gl_2$ module.
We then compute
\begin{align}
\left(A+B\right)\underline{\otimes}\left\{2\right\}&=\sum_{\mu =\left\{0\right\}, \left\{1\right\}, \left\{2\right\}} \left(A\underline{\otimes}\, \left\{2\right\}/\mu\right)\otimes \left(B\underline{\otimes} \mu\right)\nonumber\\ 
&=  \left(A\underline{\otimes} \left\{2\right\}\right)\otimes \left(B\underline{\otimes} \left\{0\right\}\right)+\left(A\underline{\otimes} \left\{1\right\}\right)\otimes \left(B\underline{\otimes} \left\{1\right\}\right)+\left(A\underline{\otimes} \left\{0\right\}\right)\otimes \left(B\underline{\otimes} \left\{2\right\}\right)\nonumber\\
&=A\underline{\otimes} \left\{2\right\}+A\otimes B+ B\underline{\otimes }\left\{2\right\},\nonumber
\end{align}
where, in the final line, we have implemented the plethysms $A\underline{\otimes}\left\{1\right\} = A$ and $B\underline{\otimes}\left\{1\right\} = B$,  and removed the trivial plethysms $A\underline{\otimes} \left\{0\right\}$ and $B\underline{\otimes} \left\{0\right\}$ (which, incidentally, correspond exactly to the modules $\mathcal{U}_0\cong \mathbb{C}$ and $\mathcal{V}_0\cong \mathbb{C}$, respectively). 
Now, considered as an $sl_2$ module, $A\underline{\otimes} \{2\}$ is irreducible with dimension 3, and similarly considered as a $sl_2\oplus l_2$ module $A\otimes B$ has dimension $2\times 2=4$ and is irreducible because $A$ is irreducible.
However, the general theory in \cite{sumner2008} establishes that $B\otimes \{2\}$ contains a one-dimensional submodule of $l_2$; hence we conclude $f_{(2)}\!=\!1$.
A similar calculation establishes
\[
\left(A+B\right)\underline{\otimes}\{1^2\}=A\underline{\otimes}\{1^2\}+A\otimes B+B\underline{\otimes}\{1^2\},
\]
and, since both $A\underline{\otimes} \{1^2\}$ and $B\underline{\otimes} \{1^2\}$ are one-dimensional $gl_2$ modules and hence also one-dimensional as $sl_2<gl_2$ and $l_2<gl_2$ modules, respectively, we find $f_{(1^2)}=2$.
\\
\noindent
\mbox{}\hfill $\Box$
\\

In ${\mathfrak S}_2$, $(2)$ is the trivial character and $(1^2)$ is the $sgn$ character. 
Hence, the ``inner'' products $(2)\ast (2) = (2)$, $(2)\ast (1^2) = (1^2)$, $(1^2)\ast (1^2) = (2)$ are completely straightforward,  and, appealing to associativity, we have
\[
g^{(2)}_{\sigma_1\sigma_2\ldots \sigma_L}:=
\left\{\text{multiplicity of }(2)\text{ in }\sigma_1\ast \sigma_2 \ast \ldots \ast \sigma_L\right\}
=
\left\{
\begin{array}{l}
1,\text{ if }\# \sigma_i=(1^2)\text{ is even},\\
0,\text{ otherwise.}
\end{array}
\right.
\]
Hence applying Lemma 4, we have
$n_2 = \sum_{\ell = 0}^{\lfloor L/2\rfloor} {L\choose 2\ell}\,2^{2\ell}$ which yields the formula given in Theorem~1 above. 
\\

\noindent 
\textbf{Lemma 6}: \textbf{Calculation of $f_\sigma$ for $\mathcal{S}$ at degree $D=3$.} \\

\noindent
At degree $D=3$ there are two partitions $\sigma\vdash D$ given by $(3)$, $(2,1)$ and $(1^3)$.
Symmetric function manipulations establish that $f_{(3)}=f_{(1^3)}=1$ and $f_{(21)}=2$.
\\

\noindent
\textbf{Proof}:
With the notation from the previous Lemma, we compute:
\begin{align}
(A+B)\underline{\otimes}\left\{3\right\} 
&= \sum_{\mu =\{0\},\{1\},\{2\},\{3\}}
\left(A\underline{\otimes}\left\{3\right\}/\left\{\mu\right\}\right)
\otimes \left(B\underline{\otimes}\left\{\mu\right\}\right)\nonumber\\
&=A\underline{\otimes}\left\{3\right\} +\left(A\underline{\otimes}\left\{2\right\}\right)\otimes B+A\otimes\left(B\underline{\otimes}\left\{2\right\}\right)+B\underline{\otimes}\left\{3\right\}\nonumber.
\end{align}
Similarly, we find:
\begin{align}
\left(A+B\right)\underline{\otimes}\left\{21\right\}&=\sum_{\mu=\left\{0\right\},\left\{1\right\},\left\{1^2\right\},\left\{2\right\},\left\{21\right\}}\left(A\underline{\otimes}\left\{21\right\}/\left\{\mu\right\}\right)
\otimes \left(B\underline{\otimes}\left\{\mu\right\}\right)\nonumber\\
&=A\underline{\otimes}\left\{21\right\}+\left(A\underline{\otimes}\{2\}\right)\otimes B+\left(A\underline{\otimes}\{1^2\}\right)\otimes B
+A\otimes\left(B\underline{\otimes}\{1^2\}\right)\nonumber\\
&\hspace{8em}+A\otimes\left(B\underline{\otimes}\{2\}\right)
+
B\underline{\otimes }\left\{21\right\}\nonumber,
\end{align}
and
\begin{align}
\left(A+B\right)\underline{\otimes}\left\{1^3\right\}&=\sum_{\mu=\left\{0\right\},\left\{1\right\},\left\{1^2\right\},\left\{1^3\right\}}\left(A\underline{\otimes}\left\{1^3\right\}/\left\{\mu\right\}\right)
\otimes \left(B\underline{\otimes}\left\{\mu\right\}\right)\nonumber\\
&=A\underline{\otimes}\left\{1^3\right\}+\left(A\underline{\otimes}\{1^2\}\right)\otimes B+A\otimes \left(B\underline{\otimes}\{1^2\}\right)
+B\underline{\otimes}\{1^3\}\nonumber.
\end{align}
We identify a single one-dimensional $sl_2\oplus l_2$ module inside each of $B\underline{\otimes}\left\{3\right\}$, $\left(A\underline{\otimes}\{1^2\}\right)\otimes B$, $B\underline{\otimes }\left\{21\right\}$, and $\left(A\underline{\otimes}\{1^2\}\right)\otimes B$. 
From this we conclude that $f_{(3)}=f_{(1^3)}=1$ and $f_{(21)}=2$, as required.
\noindent
\mbox{}\hfill $\Box$
\\

As characters of $\mathfrak{S}_3$, we have $\sigma\!=\!(3),(21)$ or $(1^3)$ and the inner products $(3)\ast \sigma=\sigma$, $(21)\ast (21)=(3)+(21)+(1^3)$, $(21)\ast (1^3)=(21)$ and $(1^3)\ast (1^3)=(3)$.
From these products it is straightforward to establish a recurrence relation for the expansion of $\sigma_1\ast \sigma_2\ast \ldots \ast \sigma_L\equiv (3)^i\ast (21)^j\ast (1^3)^k= (21)^j\ast (1^3)^k$ which shows
\[
g_{\sigma_1\sigma_2\ldots \sigma_L}^{(3)}=
\left\{
\begin{array}{l}
\frac{1}{3}(2^{k-1}-(-1)^{k-1}),\text{ if }k\neq 0, \\
\frac{1}{2}(1+(-1)^j),\text{ otherwise}.
\end{array}
\right.
\]
From this we find that the number of cubic Markov invariants $n_3$ for the strand symmetric model is given by
\begin{align}
n_3&=\sum_{\sigma_1,\sigma_2,\ldots,\sigma_L\vdash 3}g_{\sigma_1\sigma_2\ldots\sigma_L}^{(3)}f_{\sigma_1}f_{\sigma_2}\cdots f_{\sigma_L}\nonumber\\
&=\left(\sum_{k=1}^L{ L \choose k}\frac{1}{3}\left(2^{k-1}-(-1)^{k-1}\right)2^k\right)+\left(\sum_{j=0}^L{ L \choose j}\frac{1}{2}\left(1+(-1)^j\right)\right)\nonumber\\
&=6^{L-1},\nonumber
\end{align}
as claimed in Theorem 2.

\end{appendix}
%
%
%

\def\topsep{0pt}
\def\parsep{0pt plus 5pt minus 1pt}
\def\itemsep{-0.5ex}

\mbox{}
\vfill
\end{document}